\documentclass[aps,prl,twocolumn,showpacs,superscriptaddress]{revtex4}
\usepackage[dvipdf]{graphicx}
\begin{document}

\title{Observation of collapse of pseudospin order in bilayer quantum Hall ferromagnets}

\author{Stefano Luin}
\affiliation{NEST-INFM and Scuola Normale Superiore, Piazza dei
Cavalieri 7, I-56126 Pisa (Italy)}
\affiliation{Bell Labs, Lucent Technologies, Murray Hill, New
Jersey 07974}
\author{Vittorio Pellegrini}
\affiliation{NEST-INFM and Scuola Normale Superiore, Piazza dei
Cavalieri 7, I-56126 Pisa (Italy)}
\author{Aron Pinczuk}
\affiliation{Dept. of Physics, Dept. of Appl. Phys.
and Appl. Math, Columbia University, New York, New York 10027}
\affiliation{Bell Labs, Lucent Technologies, Murray Hill, New
Jersey 07974}
\author{Brian S. Dennis}
\author{Loren N. Pfeiffer}
\author{Ken W. West}
\affiliation{Bell Labs, Lucent Technologies, Murray Hill, New
Jersey 07974}

\date{\today}

\begin{abstract}
The Hartree-Fock paradigm of bilayer quantum Hall states with finite tunneling at filling factor $\nu$=1 has full pseudospin ferromagnetic order with all the electrons in the lowest symmetric Landau level. Inelastic light scattering measurements of low energy spin excitations reveal major departures from the paradigm at relatively large tunneling gaps. The results indicate the emergence of a novel correlated quantum Hall state at $\nu$=1 characterized by reduced pseudospin order. Marked anomalies occur in spin excitations when pseudospin polarization collapses by application of in-plane magnetic fields.
\end{abstract}


\maketitle
\begin{figure}[tb]
\includegraphics[width=\columnwidth]{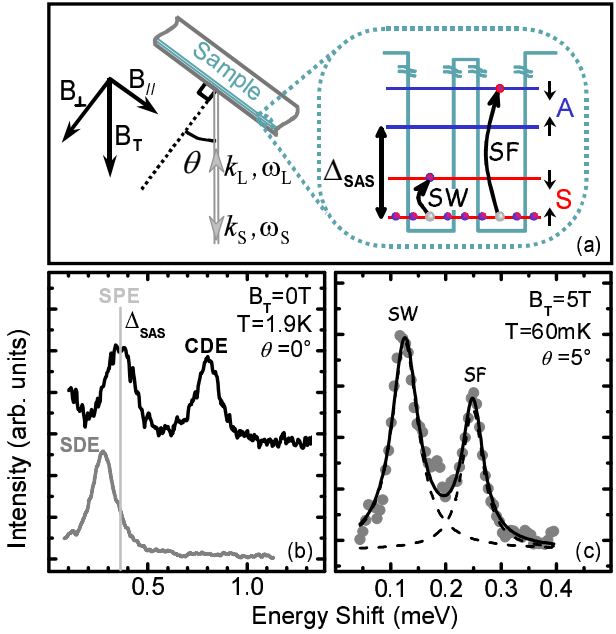}%
\caption{\label{Fig1} (a) Schematic representation of the backscattering
geometry and of the double quantum well in the single particle configuration at $\nu$=1. 
Transitions in the spin wave (SW) and spin flip (SF) modes are
indicated by curved vertical arrows; short vertical arrows indicate the
orientations of spin, S and A label the symmetric and
anti-symmetric levels separated by the tunneling gap
$\Delta_{\textrm{\tiny SAS}}$.
$k_{\textrm{\tiny L(S)}}$ and $\omega_{\textrm{\tiny L(S)}}$ indicate the wavevector and frequency
of incident (scattered) light. $B_T$, $B_{\perp}$ and $B_{/\!/}$
are the total magnetic field and its components perpendicular and
parallel to the plane of the sample. $\theta$ is the tilt angle.
(b) Polarized (black curve) and cross-polarized (gray curve)
inelastic light scattering spectra at $B_T$=0 and normal
incidence. SDE and CDE label the peak due to spin and charge
density excitations, respectively, and SPE the single particle
excitation at $\sim\! \Delta_{\textrm{\tiny SAS}}$. (c) Resonant
inelastic light scattering spectrum of SW and SF excitations at
$\nu$=1 after conventional subtraction of the background due to
the laser and main luminescence. Solid and dashed lines show the
fit with two Lorentzian functions.}
\end{figure}

%
Electron bilayers in semiconductor quantum structures embedded in a 
quantizing perpendicular magnetic field ($B_{\perp}$) are contemporary 
realizations of collective systems where bizarre quantum phases 
may appear \cite{Sar97,Pel98,Yan94,Gir97}. In particular, at Landau level filling factor
$\nu$=1 they may support excitonic-superfluidity when there is no
tunneling between the layers \cite{Gir97,Fer89,Wen92}. These
phases are driven by unique interplays between intra- and
inter-layer Coulomb interactions and have been the subject of
large research efforts in last years
\cite{Kel04,Tut04,Eis04,Ste01,Fog01,Sch01}. The bilayer system is
also characterized by the `bare', or Hartree, tunneling gap
$\Delta_{\textrm{\tiny SAS}}$ that represents the splitting
between the symmetric and anti-symmetric linear combinations of
the lowest quantum well levels, as shown in Fig.~\ref{Fig1}a. The
ground state in the presence of tunneling displays the well known
manifestations of the incompressible quantum Hall (QH) fluid
(dissipationless longitudinal transport and quantized Hall
resistance). Interactions however create intriguing behaviors such
as the disappearance at $\nu$=1 of QH signatures
when $d/l_B$ ($d$ is the inter-well distance and $l_B$ the
magnetic length) is increased above a critical value
\cite{Mur94,Boe90}.
\par
The states of electron bilayers are efficiently described by
introducing a pseudospin operator {\boldmath $\tau$}
\cite{Gir97}. Electrons in the left quantum well have pseudospin
along the $+z$ direction normal to the plane, and those in the
right well have pseudospin along $-z$. The symmetric
(anti-symmetric) states have pseudospin aligned along the $+x$
($-x$) direction. The mean field Hartree-Fock configuration of QH incompressible states with only the
lowest symmetric level populated is shown in Fig.~\ref{Fig1}a.
This state has full pseudospin polarization (all pseudospins along
the $x$-direction) with an order parameter given by the average
value of the normalized pseudospin polarization $\left\langle
\tau^x\right\rangle$=1. 

Here we report direct evidence that 
the pseudospin order of the Hartree-Fock paradigm is lost. We show that the unexpected QH states with 
reduced pseudospin polarization are probed by inelastic light scattering measurements 
of low-lying spin excitations. 
One of the excitations is the long wavelength ($q\rightarrow0$) spin-flip (SF) mode that is built with transitions across $\Delta_{\textrm{\tiny SAS}}$ with simultaneous change in spin
orientation. The other is the long wavelength spin-wave (SW) excitation built from
transitions across the Zeeman gap $E_Z$ of the lowest spin-split
symmetric levels. The transitions that build SW and SF excitation modes are depicted in Fig.~\ref{Fig1}a. 
The time-dependent Hartree-Fock approximation (TDHFA) that has been extensively 
employed to interpret bilayer experiments \cite{Luin03,Bre90,Mac90,Boe90} 
dictates that in the mean-field  $\nu$=1 state with $\left\langle
\tau^x\right\rangle$=1, the splitting between long wavelength SF and SW modes is 
$\delta E_{\tau}$=$\Delta_{\textrm{\tiny SAS}}$ \cite{Bre90}. 
We discuss below that the measured deviations of 
$\delta E_{\tau}$ from $\Delta_{\textrm{\tiny SAS}}$ provides direct evidence of
the suppression of $\left\langle \tau^x \right\rangle$. 
The pseudospin order parameter can be written as $\left\langle
\tau^x\right\rangle$=$\frac{n_S-n_{AS}}{n_S+n_{AS}}$, where $n_S$
and $n_{AS}$ are expectation values of electron densities in symmetric and
anti-symmetric levels, respectively. The reduced pseudospin order reported here thus implies that 
a new highly-correlated incompressible fluid at $\nu$=1 is formed by mixing states 
with both symmetric and antisymmetric electrons despite the large value of the tunneling gap.

Experiments were carried out in two nominally symmetric
modulation-doped Al$_{0.1}$Ga$_{0.9}$As double quantum wells (DQWs) grown by molecular
beam epitaxy and having total electron densities $n \!\sim\!
1.1$--$1.2\times10^{11}$cm$^{-2}$. Resonant inelastic light scattering
was performed on samples in a dilution refrigerator with a base
temperature of $\sim\!50$mK and with the tilted angle geometry
shown in Fig.~\ref{Fig1}a. Dye or titanium-sapphire lasers were
tuned to the fundamental optical gap of the DQW. Figure 1b
displays light scattering spectra of tunneling excitations that
illustrate the determination of $\Delta_{\textrm{\tiny SAS}}$ at
$B$=0. The spectra show collective modes in charge and spin, CDE
and SDE, and also the single-particle transition SPE \cite{Pla97}. In
the case of samples studied here static exchange and correlation
corrections at $B$=0 are small due to similar populations and
density probability profiles of the symmetric and anti-symmetric
subbands \cite{Tam94,Bol00}. We find that LDA (local density
approximation) estimates of the tunneling gap differ from Hartree
$\Delta_{\textrm{\tiny SAS}}$ values and from the measured SPE
position by less than 10$\%$ \cite{PRL04note2}. We therefore use
the measured SPE energy as the tunneling gap
$\Delta_{\textrm{\tiny SAS}}$ ($\Delta_{\textrm{\tiny
SAS}}$=0.36meV at $B$=0 for the sample shown in Fig.~\ref{Fig1}b).
At the $\nu$=1 incompressible quantum Hall states, the values of
$\frac{\Delta_{\textrm{\tiny SAS}}}{E_c}$ for the two samples are
$\sim\!0.036$ and 0.06 ($\Delta_{\textrm{\tiny SAS}}$=0.36meV and
0.58meV; $E_c=\frac{e^2}{\varepsilon l_B}$ is the average Coulomb interaction energy per electron and $\varepsilon$ is the static dielectric constant), with $\frac{d}{l_B}\!\sim\!2.2$ and 2, respectively.

\begin{figure}[tb]
\includegraphics[width=\columnwidth]{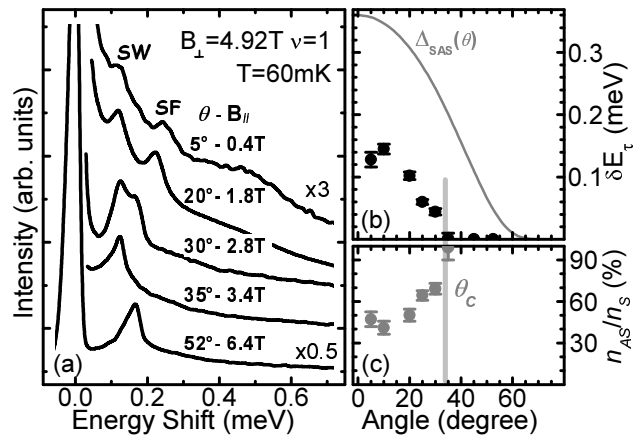}%
\caption{\label{Fig2} (a) Resonant inelastic light scattering
spectra showing spin wave (SW) and spin flip (SF) excitations as a
function of tilt angle $\theta$ at $\nu$=1. Values of the in-plane magnetic
field $B_{/\!/}$ are also shown. At
$\theta\!\geq\!\theta_c \!\approx\! 35^{\circ}$ only one peak identified as the
SW is observed. (b) Angular dependence of the energy difference
$\delta E_{\tau}$ between the SW and SF peaks (dots). The gray
curve is the predicted angular dependence of the tunneling gap
evaluated in Ref. \cite{Hu92}. (c) Angular dependence of the
emergent fraction of electrons in the excited state
$\frac{n_{AS}}{n_S}$ ($n_{AS}$ and $n_S$ are the average electron
densities in the lowest and excited levels,
respectively).}
\end{figure}

The spectrum of low-lying excitations of the sample with
$\Delta_{\textrm{\tiny SAS}}$=0.36meV at the lowest tilt angle
$\theta$=$5^{\circ}$ and $\nu$=1 is shown in Fig.~\ref{Fig1}c. Two
peaks labeled SW and SF are observed. The SW peak is the long
wavelength spin wave that occurs at $E_Z$ as required by Larmor
theorem. The higher energy peak, labeled SF, is also due to spin
excitations because it displays light scattering selection rules
identical to the SW peak. We assign the SF feature to the long
wavelength spin-flip (SF) excitation across the tunneling gap. The
identification is supported by the angular dependence displayed in
Fig.~\ref{Fig2}a, showing that the SF energy approaches $E_z$ as $\Delta_{\textrm{\tiny SAS}}$ is
reduced by the in-plane component of magnetic field $B_{/\!/}$ \cite{Hu92}.

To evaluate the impact of these results we recall again that in
TDHFA $\delta E_{\tau} $=$ \Delta_{\textrm{\tiny SAS}}$
\cite{Bre90}. At the angle of $\theta$=$5^{\circ}$, where
$B_{/\!/}$ is quite small and finite angle corrections are
negligible, we find that $\delta E_{\tau}$=$0.13\pm0.01$meV, much
smaller than the value of $\Delta_{\textrm{\tiny
SAS}}\!\sim\!0.36$meV determined from the spectra in
Fig.~\ref{Fig1}b. This result uncovers a major breakdown of the TDHFA predictions, in particular that the state has full pseudospin polarization. It is extremely important that the bilayers at $\nu$=1 continue to display well-defined magneto-transport signatures characteristics of a QH incompressible fluid \cite{Mur94, PRL04note1}. The implication is that the emergent highly-correlated fluid revealed by the light scattering measurements does not significantly change electrical conduction in the $\nu$=1 QH state at low temperatures. 

To gain more quantitative insights into the effect of correlations and pseudospin reduction on spin modes we can use the pseudospin language introduced above and the coupled spin-pseudospin bilayer Hamiltonian $\widehat{\mathcal{H}}$
derived in Ref.~\cite{Bur02}. This provides a framework to analyze the dynamics of electron spins and pseudospins in incompressible bilayers at $\nu$=1, where in-plane fluctuations of total charge density can be neglected. $\widehat{\mathcal{H}}$ is written in terms of spin and pseudospin operators ($\textbf{S}_i$
and $\textbf{T}_i = \frac{1}{2}\mbox{\boldmath $\tau$}_i$
respectively) acting on states belonging to a complete set of
localized orbital wavefunctions \{$i$\} in the lowest Landau level
\cite{Bur02}. It includes all relevant interactions in both spin and pseudospin channels, 
and coupling terms between spin and pseudospin operators. 

In order to derive the energies of long-wavelength SW and SF we note that these spin excitations 
correspond to in-phase and out-of-phase spin modes in the two layers, respectively. 
The associated states can then be constructed by using projection operators $1\pm T_i^z$ 
into the first and second layer and the spin-lowering operator $S_i^-$. Following this 
procedure the associated states can be written as
\begin{eqnarray}
\left|\Psi_{SW}\right\rangle & \!=\! &N^{-\frac{1}{2}}
\sum_iS_i^-\left|\Psi_0\right\rangle, \\
\left|\Psi_{SF}\right\rangle & \!=\! & N^{-\frac{1}{2}}\sum_i
\tau_i^z S_i^-\left|\Psi_0\right\rangle,
\end{eqnarray}
where $N$ is the total number of electrons, and $\left|\Psi_0\right\rangle$
is the ferromagnetic, fully spin polarized (at zero temperature) QH ground state with any degree of pseudospin polarization \cite{Bur02}. Using the Hamiltonian $\widehat{\mathcal{H}}$
described above, it is possible to write the energy difference $\delta 
E_{\tau}$ between SF and SW as:
\begin{equation}
\delta E_{\tau} \!=\!
\left\langle\Psi_{SF}\right| \widehat{\mathcal{H}} \left|\Psi_{SF}\right\rangle \!-\!
\left\langle\Psi_{SW}\right| \widehat{\mathcal{H}} \left|\Psi_{SW}\right\rangle \!=\!
\Delta_{\textrm{\tiny SAS}} \!\cdot\! \left\langle \tau^x\right\rangle\!, \label{dEt}
\end{equation}
where the last result follows a straightforward calculation. Eq.~(\ref{dEt}) remarks that a reduced pseudospin order parameter determines the tunneling SF energy. This conclusion is likely to remain valid even when $B_{/\!/}\!\neq\!0$. In this case phase differences are introduced between the wave functions in the two layers; their impact can be described in terms of a tumbling of the pseudo-magnetic field associated to the tunneling gap along the x-y plane. In the commensurate phase expected at relatively low $B_{/\!/}$ \cite{Mur94} pseudospins are thus aligned in different directions as a function of in-plane positions \cite{Yan94}. In
Eq.~(\ref{dEt}) therefore $\left\langle \tau^x \right\rangle$
must be replaced by the average of  {\boldmath{$\tau$}}$_i$ in the
direction of this pseudo-magnetic field. We call this quantity
$\left\langle\tau(\theta)\right\rangle$.  In the mean-field configuration, neglecting correlation effects, we continue to have $\left\langle\tau(\theta)\right\rangle$=1 and $\delta E_{\tau} \!=\!
\Delta_{\textrm{\tiny SAS}}$. 

The prediction reported in Eq.~(\ref{dEt}) allows us to link the measured SF-SW splitting 
$\delta E_{\tau}$ with a reduced $\left\langle \tau^x\right\rangle$ beyond TDHFA. We stress that
by reducing the pseudospin order electrons can efficiently optimize their inter- and intra-layer correlations by decreasing the charging energy associated to fluctuations in layer occupation (fluctuation of the pseudospin in the $z$-direction). For $\theta$=5$^{\circ}$, $\left\langle \tau^x\right\rangle\!=\! \frac{n_S-n_{AS}}{n_S+
n_{AS}}\!\approx\!0.36$, showing that the new state with reduced pseudospin order is characterized by an
high expection value $\frac{n_{AS}}{n_S+ n_{AS}}\!\approx\!0.32$ (or 32\%) for the 
fraction of the total electron density into the anti-symmetric level. It is therefore tempting to describe
this QH incompressible state in terms of bound electron-hole pairs across the tunneling gap making a particle-hole transformation in the lowest symmetric Landau level \cite{Gir84,Rez90}. 

It is surprising to observe this major loss of pseudospin ferromagnetic order 
at sizable values of  $\Delta_{\textrm{\tiny SAS}}$. Correlations in electron bilayers at $\nu=1$ were theoretically 
evaluated within models that consider the impact of quantum fluctuations from low-lying tunneling modes 
such as magneto-rotons \cite{Nak97,Jog02,Bur02,Moo97}. In these theories pseudospin order parameter 
is suppressed because of these in-plane pseudospin quantum fluctuations, leading eventually to the incompressible-compressible phase transition 
associated to the disappearing of the QH state. Because of the difference between intra- and inter-layer 
interactions, in fact, $\tau^x$ is not a good quantum number and it fluctuates in the true many-body ground state. 
In agreement with our findings, these calculations suggest that significant loss of pseudospin polarization 
could occur at high $\frac{\Delta_{\textrm{\tiny SAS}}}{E_c}$ values.

The suppression of $\tau^x$ is influenced by $\Delta_{\textrm{\tiny SAS}}$. For the sample with larger
$\Delta_{\textrm{\tiny SAS}}$=0.58meV the extrapolated value at zero angle yields  $\frac{n_{AS}}{n_S+ n_{AS}}\!\sim\!17\%$ ($\left\langle \tau^x\right\rangle\!\sim\!0.65$). We have also reduced $\Delta_{\textrm{\tiny SAS}}$ by increasing $B_{/\!/}$ at larger tilt angles $\theta$ \cite{Hu92}.
Figs.~\ref{Fig2}a and \ref{Fig2}b show that with increasing angles
$\delta E_{\tau}$ shrinks until it collapses at a `critical' angle
$\theta_c\!\sim\!35^{\circ}$. Points in Fig.~\ref{Fig2}b are the
SF-SW splitting $\delta E_{\tau}$ as a function of angle. We have
not been able to observe SF modes for angles $\theta\! >
\!\theta_c$.
Figure \ref{Fig2}c shows the ratio $\frac{n_{AS}}{n_S}\! =\!
\frac{1-\left\langle\tau(\theta)\right\rangle}{1 +
\left\langle\tau(\theta)\right\rangle}$ determined from the
measured splitting $\delta E_{\tau}(\theta)$ and from $\delta E_{\tau}(\theta)\! =\! \Delta_{\textrm{\tiny
SAS}}(\theta)\cdot\left\langle\tau(\theta)\right\rangle$, where
$\Delta_{\textrm{\tiny SAS}}(\theta)$ includes the single-particle
angular dependence derived in Ref.~\cite{Hu92} and plotted in
Fig.\ref{Fig2}b. Within this framework $\frac{n_{AS}}{n_S}\rightarrow1$ in a continuous way ($\left\langle\tau(\theta)\right\rangle\rightarrow0$) when $\theta\rightarrow\theta_c$. It is possible, however, that close to the collapse of $\delta E_{\tau}$, higher-order corrections to Eq.~(\ref{dEt}) will affect the precise determination of the pseudospin polarization.

The data, however, reveal strong increase of correlations as $\Delta_{\textrm{\tiny SAS}}(\theta)$
diminishes. The value of $\Delta_{\textrm{\tiny SAS}}(\theta_c)$ is consistent with the phase transformation to the compressible phase (without QH effect) \cite{Gir97,Mur94,Yan94}. Given the values in our samples of $\frac{\Delta_{\textrm{\tiny SAS}}}{E_c}$ and $\frac{d}{l_B}$ the commensurate-incommensurate phase transition observed in Ref.~\cite{Mur94} is not expected to occur here for $\theta \!\leq\! \theta_c$. Additionally, the transition to the incommensurate phase would produce a non-observed abrupt reduction of $\left\langle \tau_x (\theta)\right\rangle$. It is possible, however, that the decrease of exchange energy associated to the effect of $B_{/\!/}$ could contribute to increase quantum fluctuations, further reducing the order parameter.
\begin{figure}[!bt]
\includegraphics[width=\columnwidth]{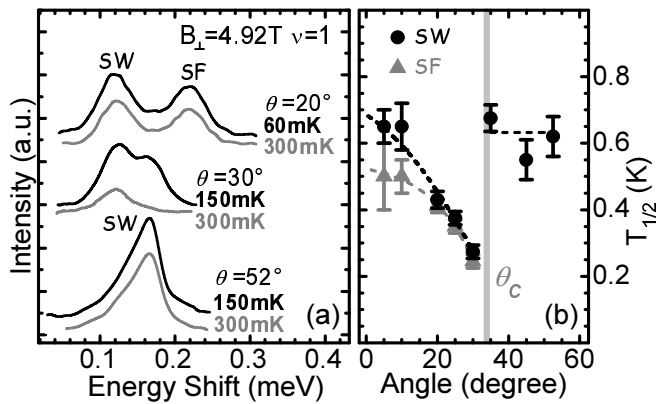}%
\caption{\label{Fig3}(a) Resonant inelastic light scattering
spectra of spin wave (SW) and spin flip (SF) excitations at $\nu$=1 and
different values of temperature and tilt angles $\theta$, after
conventional background subtraction. (b) Angular dependence of the
temperature $T_{1/2}$. $T_{1/2}$ is defined as the temperature at
which the intensity of the SW (black dots) and SF (grey triangles)
peaks is reduced to half of the lowest temperature value. The
dashed lines are guides for the eyes. $\theta_c$ is the angle at
which the SW and SF peak merge.}
\end{figure}

Evidence of a phase transition at the collapse of pseudospin order at  $\theta_c$ is also seen in the marked temperature dependence of SF and SW modes, as shown in Fig.~\ref{Fig3}a. For typical spectra
with $\theta \!<\! \theta_c$ the intensities of the peaks have a large temperature dependence. The temperature dependence is slower for the asymmetric SW modes measured at $\theta\!>\!\theta_c$. To describe this behavior
we introduce a temperature $T_{1/2}$ at which the peak intensity is half of its value at the lowest temperature. Figure~\ref{Fig3}b displays $T_{1/2}$ versus angle for both SW (black circles) and SF (gray triangles). It can be seen that at $\theta_c$ there is an abrupt change in $T_{1/2}$ that accompanies the
disappearance of the SF mode. The strong decrease of $T_{1/2}$ for the SW for $\theta\!<\!\theta_c$ suggests a rich spin dynamics in the ferromagnetic $\nu=1$ state that may be linked to
spin-pseudospin coupling. 

In conclusion we determined major reductions of pseudospin
ferromagnetic order due to correlations in the incompressible phase of coupled bilayers at
$\nu$=1 by inelastic light scattering measurements of spin
excitations. The results are surprising by
revealing large correlation effects in a range of
$\frac{\Delta_{\textrm{\tiny SAS}}}{E_c}$ and $\frac{d}{l_B}$
values where magneto-transport results find well-defined QH
signatures. Further studies, including absorption across the fundamental gap between valence and
conduction bands \cite{Man00,Cot01}, should clarify if the correlated state with reduction of 
pseudospin order displays the properties of an electron-hole excitonic quantum fluid.  

\begin{acknowledgments}
VP acknowledges support from the Italian Ministry of Foreign
Affairs, Italian Ministry of Research and from the European
Community's Human Potential Program (Project HPRN-CT-2002-00291).
AP acknowledges the National Science Foundation under Award Number
DMR-03-52738, the Department of Energy under award
DE-AIO2-04ER46133, and the W. M. Keck Foundation. We thank Allan H. MacDonald for illuminating
discussions.
\end{acknowledgments}

\bibliography{PRLuin04}

\begin{thebibliography}{31}
\bibitem[{PRL({\natexlab{b}})}]{PRL04note1}
\bibinfo{note}{Magneto-transport experiments down to 300mK show the
  manifestations of a quantum Hall state at $\nu$=1 at angles below
  30$^{\circ}$. Detailed temperature dependence studies are hindered by
  parallel conduction.}
\end{thebibliography}
\end{document}